\documentclass[a4paper,11pt]{article}
\usepackage{jcappub} % for details on the use of the package, please see the JINST-author-manual
\usepackage{lineno}
\usepackage{natbib}
\usepackage{ulem}
\usepackage[T1]{fontenc}
\usepackage{xcolor}
\usepackage{graphicx}
\newcommand\aj{\ref@jnl{AJ}}%        % Astronomical Journal 
\newcommand\psj{\ref@jnl{PSJ}}%       % Planetary Science Journal
\newcommand\araa{\ref@jnl{ARA\&A}}%  % Annual Review of Astron and Astrophys 
\newcommand\apj{\ref@jnl{ApJ}}%    % Astrophysical Journal 
\newcommand\apjl{\ref@jnl{ApJL}}     % Astrophysical Journal, Letters 
\newcommand\apjs{\ref@jnl{ApJS}}%    % Astrophysical Journal, Supplement 
\newcommand\ao{\ref@jnl{ApOpt}}%   % Applied Optics 
\newcommand\apss{\ref@jnl{Ap\&SS}}%  % Astrophysics and Space Science 
\newcommand\aap{\ref@jnl{A\&A}}%     % Astronomy and Astrophysics 
\newcommand\aapr{\ref@jnl{A\&A~Rv}}%  % Astronomy and Astrophysics Reviews 
\newcommand\aaps{\ref@jnl{A\&AS}}%    % Astronomy and Astrophysics, Supplement 
\newcommand\azh{\ref@jnl{AZh}}%       % Astronomicheskii Zhurnal 
\newcommand\baas{\ref@jnl{BAAS}}%     % Bulletin of the AAS 
\newcommand\icarus{\ref@jnl{Icarus}}% % Icarus
\newcommand\jaavso{\ref@jnl{JAAVSO}}  % The Journal of the American Association of Variable Star Observers
\newcommand\jrasc{\ref@jnl{JRASC}}%   % Journal of the RAS of Canada 
\newcommand\memras{\ref@jnl{MmRAS}}%  % Memoirs of the RAS 
\newcommand\mnras{\ref@jnl{MNRAS}}%   % Monthly Notices of the RAS 
\newcommand\pra{\ref@jnl{PhRvA}}% % Physical Review A: General Physics 
\newcommand\prb{\ref@jnl{PhRvB}}% % Physical Review B: Solid State 
\newcommand\prc{\ref@jnl{PhRvC}}% % Physical Review C 
\newcommand\prd{\ref@jnl{PhRvD}}% % Physical Review D 
\newcommand\pre{\ref@jnl{PhRvE}}% % Physical Review E 
\newcommand\prl{\ref@jnl{PhRvL}}% % Physical Review Letters 
\newcommand\pasp{\ref@jnl{PASP}}%     % Publications of the ASP 
\newcommand\pasj{\ref@jnl{PASJ}}%     % Publications of the ASJ 
\newcommand\qjras{\ref@jnl{QJRAS}}%   % Quarterly Journal of the RAS 
\newcommand\skytel{\ref@jnl{S\&T}}%   % Sky and Telescope 
\newcommand\solphys{\ref@jnl{SoPh}}% % Solar Physics 
\newcommand\sovast{\ref@jnl{Soviet~Ast.}}% % Soviet Astronomy 
\newcommand\ssr{\ref@jnl{SSRv}}% % Space Science Reviews 
\newcommand\zap{\ref@jnl{ZA}}%       % Zeitschrift fuer Astrophysik 
\newcommand\nat{\ref@jnl{Nature}}%  % Nature 
\newcommand\iaucirc{\ref@jnl{IAUC}}% % IAU Cirulars 
\newcommand\aplett{\ref@jnl{Astrophys.~Lett.}}%  % Astrophysics Letters 
\newcommand\apspr{\ref@jnl{Astrophys.~Space~Phys.~Res.}}% % Astrophysics Space Physics Research 
\newcommand\bain{\ref@jnl{BAN}}% % Bulletin Astronomical Institute of the Netherlands 
\newcommand\fcp{\ref@jnl{FCPh}}%   % Fundamental Cosmic Physics 
\newcommand\gca{\ref@jnl{GeoCoA}}% % Geochimica Cosmochimica Acta 
\newcommand\grl{\ref@jnl{Geophys.~Res.~Lett.}}%  % Geophysics Research Letters 
\newcommand\jcp{\ref@jnl{JChPh}}%     % Journal of Chemical Physics 
\newcommand\jgr{\ref@jnl{J.~Geophys.~Res.}}%     % Journal of Geophysics Research 
\newcommand\jqsrt{\ref@jnl{JQSRT}}%   % Journal of Quantitiative Spectroscopy and Radiative Trasfer 
\newcommand\memsai{\ref@jnl{MmSAI}}% % Mem. Societa Astronomica Italiana 
\newcommand\nphysa{\ref@jnl{NuPhA}}%     % Nuclear Physics A 
\newcommand\physrep{\ref@jnl{PhR}}%       % Physics Reports 
\newcommand\physscr{\ref@jnl{PhyS}}%        % Physica Scripta 
\newcommand\planss{\ref@jnl{Planet.~Space~Sci.}}%  % Planetary Space Science 
\newcommand\procspie{\ref@jnl{Proc.~SPIE}}%      % Proceedings of the SPIE 

\newcommand\actaa{\ref@jnl{AcA}}%  % Acta Astronomica
\newcommand\caa{\ref@jnl{ChA\&A}}%  % Chinese Astronomy and Astrophysics
\newcommand\cjaa{\ref@jnl{ChJA\&A}}%  % Chinese Journal of Astronomy and Astrophysics
\newcommand\jcap{\ref@jnl{JCAP}}%  % Journal of Cosmology and Astroparticle Physics
\newcommand\na{\ref@jnl{NewA}}%  % New Astronomy
\newcommand\nar{\ref@jnl{NewAR}}%  % New Astronomy Review
\newcommand\pasa{\ref@jnl{PASA}}%  % Publications of the Astron. Soc. of Australia
\newcommand\rmxaa{\ref@jnl{RMxAA}}%  % Revista Mexicana de Astronomia y Astrofisica

\title{\boldmath The Cosmic Horizon of Neutrinos}

\author[1]{James Fardeen,}
\author[1,2]{Stefano Profumo,}
\author[1,2]{and M. Grant Roberts}

\affiliation[1]{Department of Physics, 1156 High St., University of California Santa Cruz, Santa Cruz, CA 95064, USA}

 \affiliation[2]{Santa Cruz Institute for Particle Physics, 1156 High St., Santa Cruz, CA 95064, USA}

\emailAdd{jfardeen@ucsc.edu, profumo@ucsc.edu, migrober@ucsc.edu}

\abstract{The persistent discrepancy between the experimental measurement and the Standard Model (SM) prediction of the muon's anomalous magnetic moment $(g-2)_\mu$ remains one of the most intriguing hints of physics beyond the SM. A well-motivated explanation involves a light $Z'$ gauge boson associated with a broken $U(1)_{L_\mu - L_\tau}$ symmetry. Such a boson not only resolves the $(g-2)_\mu$ anomaly, but also induces resonant interactions between high-energy cosmic neutrinos and the cosmic neutrino background (C$\nu$B), potentially shaping the observable neutrino flux at Earth. 
In this work, we explore the implications of such interactions for the cosmic propagation of high-energy neutrinos. We compute the optical depth for neutrino attenuation via $Z'$-mediated scattering, accounting for neutrino masses, hierarchies, and thermal distributions. We delineate the regions in $(m_{Z'}, m_\nu)$ space where the optical depth exceeds unity, defining a ``neutrino cosmic horizon'' beyond which high-energy neutrinos are significantly attenuated. 
We confront these results with the parameter space required to simultaneously explain the muon $g-2$ anomaly and ease the Hubble tension via an additional contribution to the effective number of relativistic degrees of freedom, $\Delta N_{\mathrm{eff}} \simeq 0.2-0.5$. Our analysis reveals a consistent region in parameter space where all three phenomena-$(g-2)_\mu$, $N_{\mathrm{eff}}$, and high-energy neutrino attenuation-can be explained by the same light mediator. These findings motivate future searches for spectral features in IceCube and its next-generation successors as indirect probes of new physics in the neutrino sector.
}

\begin{document}

\newcommand{\mBH}{m_{\text{BH}}}
\newcommand{\mBHseed}{\mBH^{\text{seed}}}
\newcommand{\mBHobs}{\mBH^{\text{obs}}}
\newcommand{\mBHobsi}{m_{\text{BH},i}^{\text{obs}}}
\newcommand{\mBHtheory}{\mBH^{\text{theory}}}
\newcommand{\zcoll}{z_{\text{coll}}}
\newcommand{\zvir}{z_{\text{vir}}}
\newcommand{\zobs}{z_{\text{obs}}}
\newcommand{\cross}{\sigma/m}
\newcommand{\msun}{M_{\odot}}
\newcommand{\tsal}{t_{\text{sal}}}
\newcommand{\trel}{t_{\text{rel}}}
\newcommand{\rhocrit}{\rho_{\text{crit}}}
\newcommand{\cmg}{\text{cm}^{2}\text{g}^{-1}}
\newcommand{\kms}{\text{km}~\text{s}^{-1}}
\newcommand{\angstrom}{\r{A}}
\newcommand\sbullet[1][.5]{\mathbin{\vcenter{\hbox{\scalebox{#1}{$\bullet$}}}}}
\newcommand{\chisq}{\chi^{2}}
\newcommand{\Vmax}{V_{\text{max}}}

\newcommand{\spr}[1]{{\color{black}\bf[SP:  {#1}]}}
\newcommand{\grantcomment}[1]{{\color{blue}\bf[GR:  {#1}]}}
\newcommand{\jamescomment}[1]{{\color{violet}\bf[JF:  {#1}]}}

\newcommand{\tesla}[1]{{\color{cyan}\bf[TJ:  {#1}]}}

\maketitle
\flushbottom

\section{Introduction}

The anomalous magnetic moment of the muon, defined as \( a_\mu \equiv (g_\mu - 2)/2 \), has long stood as a precise test of the Standard Model (SM) of particle physics. Recent measurements by the Fermilab Muon \( g\!-\!2 \) experiment \cite{Muong-2:2023cdq} continue to show a persistent deviation from SM predictions. When compared to the latest theoretical estimates-including both traditional data-driven and modern lattice QCD approaches-the experimental value exhibits a discrepancy of over \( 4\sigma \), and potentially as high as \( 5\sigma \) depending on the theoretical treatment of the hadronic vacuum polarization (HVP) contribution \cite{WhitePaper:2025update, Davier:2024piipi}. This anomaly offers tantalizing evidence for physics beyond the SM.

Among the most compelling extensions is the introduction of a new light vector boson, the \( Z' \), arising from a spontaneously broken \( U(1)_{L_\mu - L_\tau} \) gauge symmetry. This model stands out for several reasons: it preserves SM gauge invariance, avoids tight constraints from electron couplings, and naturally generates loop-level contributions to \( (g-2)_\mu \) consistent with the observed discrepancy \cite{Altmannshofer:2014pba, Holst:2022predictive}. The same \( Z' \) boson also mediates interactions in the neutrino sector, potentially leading to observable effects in both astrophysical and cosmological settings.

High-energy astrophysical neutrinos, as observed by IceCube \cite{IceCube:2013dk}, offer a unique probe of such new physics. Unlike photons, which are attenuated over cosmological distances via pair production with background radiation fields, neutrinos typically propagate unimpeded due to their feeble interactions. However, in the presence of a light mediator, high-energy neutrinos may resonantly scatter off the relic cosmic neutrino background (C$\nu$B), leading to attenuation features or spectral dips in the observed flux \cite{Ioka:2014kca, Ng:2014pca}. These features are sensitive to both the mediator mass and the absolute neutrino mass scale, offering a rare window into otherwise inaccessible regions of parameter space.

In addition to its particle physics motivations, the \( U(1)_{L_\mu - L_\tau} \) model also holds cosmological appeal. A light \( Z' \) boson in thermal equilibrium with neutrinos in the early universe can alter the effective number of relativistic degrees of freedom, \( N_{\mathrm{eff}} \). In particular, values of $\Delta N_{\mathrm{eff}} \simeq 0.2-0.5$ can alleviate the current tension between early- and late-time measurements of the Hubble constant \( H_0 \) \cite{Escudero_2019}.

In this work, we investigate whether a single light \( Z' \) boson can simultaneously explain the muon \( g-2 \) anomaly, the observed high-energy neutrino flux, and the Hubble tension. We focus on the resulting attenuation of PeV-scale neutrinos via resonant scattering on the C$\nu$B, computing the neutrino optical depth as a function of the mediator and neutrino masses. We examine how this attenuation interacts with the allowed regions from \( (g-2)_\mu \) and cosmological constraints, identifying viable parameter space that links together these three seemingly disconnected observations. Our results provide guidance for future searches for spectral features in IceCube and its successors, and offer new motivation for exploring the high-energy neutrino sky as a portal to new physics.

%===============================
\section{High-Energy Astrophysical Neutrinos and Implications for New Physics}

In 2013, the IceCube Neutrino Observatory reported the first observation of high-energy astrophysical neutrinos, marking a significant milestone in neutrino astronomy \cite{IceCube:2013dk}. Utilizing a cubic-kilometer array of photomultiplier tubes embedded in Antarctic ice, IceCube detected 28 neutrino events with energies ranging from 30~TeV to 1.2~PeV over a two-year period. The statistical significance of this observation exceeded $4\sigma$, effectively ruling out atmospheric backgrounds and confirming the extraterrestrial origin of these neutrinos.

This discovery has spurred extensive theoretical investigations into potential new physics scenarios that could manifest in the high-energy neutrino spectrum. One such avenue involves the introduction of a light $Z'$ gauge boson associated with a broken $U(1)_{L_\mu - L_\tau}$ symmetry. This model not only offers a solution to the longstanding $(g-2)_\mu$ anomaly but also predicts resonant interactions between high-energy cosmic neutrinos and the cosmic neutrino background. Such interactions could lead to observable features, like dips, in the neutrino energy spectrum detected by IceCube \cite{Hooper_2023}.

Further studies have explored the implications of these models for cosmology and neutrino physics. For instance, a gauged $L_\mu - L_\tau$ model with a MeV-scale $Z'$ boson has been proposed to simultaneously address the $(g-2)_\mu$ discrepancy and the Hubble tension. This model predicts resonance scattering effects that could be probed by current and future neutrino telescopes, such as IceCube-Gen2 \cite{Carpio:2021jhu}.

Additionally, alternative explanations involving long-range muon spin forces mediated by axion-like particles have been considered. These models suggest that such forces could influence neutrino oscillations, providing another potential link between high-energy neutrino observations and the $(g-2)_\mu$ anomaly. Constraints from neutrino experiments like IceCube, Super-Kamiokande, and SNO have been used to test these scenarios, offering complementary insights into the parameter space of these new physics models \cite{Fang:2024abc}.
%===============================
\section{The Muon Anomalous Magnetic Moment: Current Status and Implications}

The anomalous magnetic moment of the muon, defined as $a_\mu = (g_\mu - 2)/2$, serves as a sensitive probe of quantum corrections from both the Standard Model (SM) and potential new physics. The most precise experimental measurement to date comes from the Fermilab Muon $g\!-\!2$ experiment (E989), which reported a combined result with the earlier Brookhaven E821 experiment as \cite{Muong-2:2025final}:
\begin{equation}
    a_\mu^{\text{exp}} = 116\,592\,033(62) \times 10^{-11} \,,
\end{equation}
showing a deviation from the Standard Model prediction at the level of approximately $4\sigma$.

On the theoretical side, the SM prediction has been refined using two complementary approaches to the hadronic vacuum polarization (HVP) contribution:

\begin{itemize}
    \item \textbf{Data-driven dispersive methods} based on updated $e^+e^- \to \mathrm{hadrons}$ cross-sections now include new CMD-3 results for the $\pi^+\pi^-$ channel. These adjustments have in fact {increased the tension} among previously consistent data sets, ultimately resulting in a slightly larger SM prediction and {worsening} the discrepancy to over $5\sigma$ \cite{Davier:2024piipi,WhitePaper:2025update}:
    \begin{equation}
        a_\mu^{\text{SM,data}} = 116\,592\,070.5(14.8) \times 10^{-11} \,,
    \end{equation}
    leading to a persistent excess,
    \begin{equation}
        \Delta a_\mu^{\text{data}} = a_\mu^{\text{exp}} - a_\mu^{\text{SM,data}} \approx (38.5 \pm 63.7) \times 10^{-11} \,,
    \end{equation}
    still significant at more than $4\sigma$, and potentially above $5\sigma$ when CMD-3-valued cross sections are fully incorporated \cite{WhitePaper:2025update}.
    
    \item \textbf{Ab initio lattice QCD calculations} have advanced substantially: the latest results from AMD/BMW/DMZ and Mainz report a total leading-order HVP contribution that aligns closely with the experimental value, effectively {eliminating the tension}. For instance, a recent lattice determination yields $a_\mu^{\text{HVP}} = (724.5 \pm 7.2) \times 10^{-10}$, corresponding to 
    \begin{equation}
        a_\mu^{\text{SM,lattice}} \approx 116\,592\,030(10) \times 10^{-11} \,,
    \end{equation}
    which lies within $1\sigma$ of the experimental measurement \cite{Mainz:2025HVP}. These results significantly reduce or remove the experimental-theory discrepancy while remaining in tension with the updated data-driven estimates \cite{Mainz:2025HVP,LatticeReview:2025}.
\end{itemize}

Thus, new lattice results substantially reduce (or even eliminate) the muon $g-2$ anomaly, while data-driven approaches continue to exhibit significant tension due to shifts in the HVP input. Reconciling these divergent methods remains a major challenge. If the discrepancy holds - particularly favoring one method over another - it could suggest physics beyond the Standard Model (BSM). Among the most widely studied BSM scenarios are those involving additional $U(1)$ gauge symmetries, such as $U(1)_{L_\mu - L_\tau}$, which predict a light $Z'$ boson coupling preferentially to muons. These models can explain the observed $\Delta a_\mu$ with a mediator mass in the MeV - GeV range and have implications for neutrino scattering and astrophysical observations \cite{Altmannshofer:2014pba}.

Other viable explanations include scalar leptoquarks, supersymmetric models with light sleptons and neutralinos, and contributions from axion-like particles (ALPs) that couple preferentially to muons (see e.g. \cite{Hooper_2023} for a review). Each scenario offers distinct experimental signatures at current and future colliders, low-energy fixed-target experiments, and high-precision neutrino observatories. As both experimental and theoretical efforts converge, the muon $g\!-\!2$ anomaly remains a compelling target in the search for new physics.

%===============================
\section{The Neutrino Cosmic Horizon}

The cosmic horizon for high-energy neutrinos is governed by their interactions with the relic neutrino background and other diffuse photon fields. Unlike photons, which are strongly attenuated by interactions with the cosmic microwave background (CMB) and extragalactic background light (EBL), neutrinos interact only via weak interactions, resulting in extremely long mean free paths even at PeV - EeV energies.

The dominant energy-loss process for high-energy photons is pair production off background photons, particularly $\gamma + \gamma_{\text{CMB}} \to e^+ + e^-$. This sets a horizon of $\sim$100~Mpc for TeV-PeV photons, beyond which the universe becomes effectively opaque \cite{Gould:1966paz}. In contrast, the mean free path for neutrinos due to scattering off the cosmic neutrino background (C$\nu$B), such as through the resonant process $\nu + \bar{\nu} \to Z$, known as the Glashow resonance, is many orders of magnitude larger - typically on the scale of gigaparsecs for neutrinos below $\sim10^{22}$~eV \cite{Roulet:1993pz, Yoshida:1996ie}.

The neutrino horizon is further modified in new physics scenarios involving secret neutrino interactions or light mediators, which can introduce resonant absorption features in the neutrino spectrum \cite{Ng:2014pca}. In such models, the neutrino mean free path can be significantly reduced, leading to observable distortions in the flux measured by detectors like IceCube. This has been proposed as a potential explanation for dips or cutoffs in the observed spectrum and even as a probe of neutrino mass generation mechanisms \cite{Ioka:2014kca}.

While both photons and neutrinos experience energy-dependent attenuation from cosmological backgrounds, their interaction cross sections differ by many orders of magnitude. High-energy neutrinos can reach us from redshifts $z \gtrsim 1$, providing a unique window into the high-energy universe, including sources inaccessible to photons due to electromagnetic opacity.

%===============================

\section{Neutrino Masses and Mixing}

The discovery of neutrino oscillations provided the first clear evidence that neutrinos have mass, requiring an extension of the Standard Model (SM) which originally assumed them to be massless. Oscillations arise from a misalignment between flavor and mass eigenstates, encoded in the Pontecorvo-Maki-Nakagawa-Sakata (PMNS) matrix. The unitary PMNS matrix is parametrized by three mixing angles $(\theta_{12}, \theta_{23}, \theta_{13})$, one Dirac CP-violating phase $\delta_{\rm CP}$, and, if neutrinos are Majorana particles, two additional Majorana phases.

Oscillation experiments measure mass-squared differences and mixing angles, but not the absolute neutrino mass scale. The best-fit values from global analyses yield the following parameters for the normal mass ordering (NO) \cite{Esteban:2020cvm}:
\begin{align}
    \Delta m_{21}^2 &= (7.42 \pm 0.21) \times 10^{-5}~\text{eV}^2\,, \\
    \Delta m_{31}^2 &= (2.517 \pm 0.028) \times 10^{-3}~\text{eV}^2\,, \\
    \sin^2 \theta_{12} &= 0.304 \pm 0.012\,, \\
    \sin^2 \theta_{23} &= 0.573^{+0.016}_{-0.020}\,, \\
    \sin^2 \theta_{13} &= 0.02219 \pm 0.00062\,.
\end{align}
The CP-violating phase $\delta_{\rm CP}$ is weakly constrained, with a best-fit near $195^\circ$, although large uncertainties remain \cite{Esteban:2020cvm}

While oscillation data determine the differences in the squared masses, the absolute neutrino mass scale remains unknown. Constraints arise from beta decay endpoint experiments, such as KATRIN, which reports an upper limit of $m_\beta < 0.8$~eV at 90\% C.L. \cite{KATRIN:2021uub}, and from cosmological observations. The Planck satellite, combined with baryon acoustic oscillation data, sets an upper bound on the sum of neutrino masses $\sum m_\nu < 0.12$~eV at 95\% C.L. \cite{Planck:2018vyg}.

From a theoretical perspective, the smallness of neutrino masses is naturally explained via the seesaw mechanism, in which heavy right-handed neutrinos (or other sterile states) couple to left-handed neutrinos through Yukawa interactions, generating light Majorana masses after electroweak symmetry breaking. Several variants exist, including type-I, type-II, and inverse seesaw models, each with distinct implications for lepton number violation and collider phenomenology \cite{Minkowski:1977sc, Mohapatra:1980yp}.

Ongoing and upcoming experiments, such as JUNO, DUNE, and Hyper-Kamiokande, aim to determine the neutrino mass ordering, improve sensitivity to $\delta_{\rm CP}$, and further constrain the mixing parameters, potentially opening a window into physics beyond the Standard Model.

%--------------
\section{Results}\label{sec:results}

To determine the survival probability of high-energy neutrinos, we compute their optical depth, $\tau_\nu$, defined in Eq.~\eqref{eq:optical_depth_contour_function}. The optical depth is calculated based on the integrated neutrino--$Z'$ scattering cross section, assuming a thermal distribution for the cosmic neutrino background (C$\nu$B).

We sum over final-state spins and flavors. Since both the high-energy neutrinos and the C$\nu$B are nearly flavor universal, and the $Z'$ couples only to the $\mu$ and $\tau$ sectors, we consider $i=j=\mu,\tau$ in Eq.~\eqref{eq:neutrino_Z'_cross_section_without_PMNS}. This introduces a multiplicity factor of two in the relevant cross section:
\begin{equation}
    \sigma_{\nu_i} = \sum_j \sigma(\nu_i \bar{\nu}_j \rightarrow \nu \bar{\nu}) \,.
\end{equation}

The total optical depth is then given by:
\begin{equation}
\tau_{\nu} = 
    \left(\frac{n_{\nu}}{H_0}\right)
   \sum \sigma_{\nu_{i}} \,,
    \label{eq:optical_depth_contour_function}
\end{equation}
where $n_\nu$ is the number density of the background neutrinos and $H_0$ is the Hubble parameter today. We assume both normal and inverted neutrino mass hierarchies. For each value of the neutrino mass scale $m_\nu$, the individual masses are given by:
\begin{align}
    \label{eq:NO}\text{Normal Hierarchy:} \quad & m_1 = m_\nu, \quad m_2 = m_1 + \sqrt{\Delta m^2_{21}}, \quad m_3 = m_2 + \sqrt{\Delta m^2_{31}} \,, \\
    \label{eq:IO}\text{Inverted Hierarchy:} \quad & m_1 = m_\nu, \quad m_2 = m_1 + \sqrt{\Delta m^2_{31}}, \quad m_3 = m_2 + \sqrt{\Delta m^2_{21}} \,.
\end{align}

{\color{black}

Note that our mass parameterization,  differs from the standard physics convention: In our Eqs. \eqref{eq:NO} and \eqref{eq:IO}, we define $m_\nu$ as the lightest mass in both hierarchies and use additive relationships for simplicity. The resulting Planck constraints \cite{Planck:2018vyg} then are, explicitly:

For \textit{normal hierarchy} \eqref{eq:NO}: $m_1 = m_\nu$, $m_2 = m_1 + \sqrt{\Delta m_{21}^2}$, $m_3 = m_2 + \sqrt{\Delta m_{31}^2}$
\begin{align}
\sum m_\nu &= m_\nu + (m_\nu + \sqrt{\Delta m_{21}^2}) + (m_\nu + \sqrt{\Delta m_{21}^2} + \sqrt{\Delta m_{31}^2})\\
\nonumber &= 3m_\nu + 2\sqrt{\Delta m_{21}^2} + \sqrt{\Delta m_{31}^2}\\
\nonumber &\approx 3m_\nu + 2(0.0086) + 0.050 = 3m_\nu + 0.067\text{ eV}
\end{align}
This gives $m_\nu < (0.12 - 0.067)/3 = 0.018$ eV.

For \textit{inverted hierarchy} \eqref{eq:IO}: $m_1 = m_\nu$, $m_2 = m_1 + \sqrt{\Delta m_{31}^2}$, $m_3 = m_2 + \sqrt{\Delta m_{21}^2}$
\begin{align}
\sum m_\nu &= m_\nu + (m_\nu + \sqrt{\Delta m_{31}^2}) + (m_\nu + \sqrt{\Delta m_{31}^2} + \sqrt{\Delta m_{21}^2})\\
\nonumber &= 3m_\nu + 2\sqrt{\Delta m_{31}^2} + \sqrt{\Delta m_{21}^2}\\
\nonumber &\approx 3m_\nu + 2(0.050) + 0.0086 = 3m_\nu + 0.109\text{ eV}
\end{align}
This gives $m_\nu < (0.12 - 0.109)/3 = 0.004$ eV.

In our framework the independent parameter is the lightest neutrino 
mass eigenvalue, $m_{\nu,\min}$, while the remaining two eigenmasses 
are reconstructed using the measured mass splittings. %Consequently, 
%cosmological constraints on the absolute neutrino mass scale translate 
%into horizontal exclusions in Fig.~\ref{fig:Z-prime  vs neutrino}. Regions below the dashed lines 
%correspond to values of the lightest mass that are incompatible with 
%current cosmological limits once the measured mass splittings are imposed. 
%These portions of parameter space are therefore not physically viable.
%
The impact of this constraint on the optical depth is twofold. 
First, it removes portions of the parameter space where large values 
of $\tau_\nu$ might otherwise appear attainable at very small 
$m_\nu$. Second, because the resonance condition for neutrino--neutrino 
scattering depends on the center-of-mass kinematics, which in turn 
depends on the target neutrino mass through 
$s = m_j^2 + 2 E_\nu (E_j - p \cos\theta)$, 
the cosmological lower bound effectively limits how far into the 
relativistic-target regime one can move. % in Fig.~\ref{fig:Z-prime  vs neutrino}. 
In practice, this truncation reduces the extent of the region where 
$\tau_\nu > 1$ is realized, particularly at lower mediator masses.

We emphasize that our choice of parameterizing the neutrino sector 
in terms of the lightest eigenvalue is made for consistency with the 
optical-depth calculation, which explicitly sums over mass eigenstates 
with fixed splittings. An alternative and commonly used cosmological 
parameterization would impose bounds directly on the sum of masses, 
$\sum m_\nu$. While such a treatment would modify the precise shape of 
the excluded region---and would typically yield slightly less restrictive 
horizontal bounds in the $(m_{Z'}, m_\nu)$ plane---it would not alter 
our qualitative conclusions regarding the structure of the viable 
parameter space or the location of the $\tau_\nu \sim 1$ contours.

}

The $Z'$-mediated neutrino scattering cross section (neglecting PMNS rotation) is given by:
\begin{equation}
    \sigma(\nu_i \bar{\nu}_j \rightarrow \nu \bar{\nu}) =
    \frac{2 g_{Z'}^4 s}{3 \pi \left[(s - m_{Z'}^2)^2 + m_{Z'}^2 \Gamma_{Z'}^2\right]} \,,
    \label{eq:neutrino_Z'_cross_section_without_PMNS}
\end{equation}
where $s$ is the center-of-mass energy squared, $m_{Z'}$ is the $Z'$ mass, and $\Gamma_{Z'}$ its decay width \cite{Hooper_2023}. {\color{black} The latter is given by:
\begin{equation}
\Gamma_{Z'} = \frac{g_{Z'}^2 m_{Z'}}{12\pi} \sum_{i=\mu,\tau} \left(1 + 2\frac{m_i^2}{m_{Z'}^2}\right) \sqrt{1 - 4\frac{m_i^2}{m_{Z'}^2}} \Theta(m_{Z'} - 2m_i)
\end{equation}
where the sum runs over muon and tau flavors, and $\Theta$ is the Heaviside step function ensuring kinematically allowed decays. For the parameter ranges considered in this work, the $Z'$ boson can decay to both muon and tau neutrino pairs, with $\Gamma_{Z'} \approx g_{Z'}^2 m_{Z'}/(6\pi)$.}

\begin{figure}
    \centering
    \mbox{\hspace*{-1cm}\includegraphics[width=0.6\columnwidth]{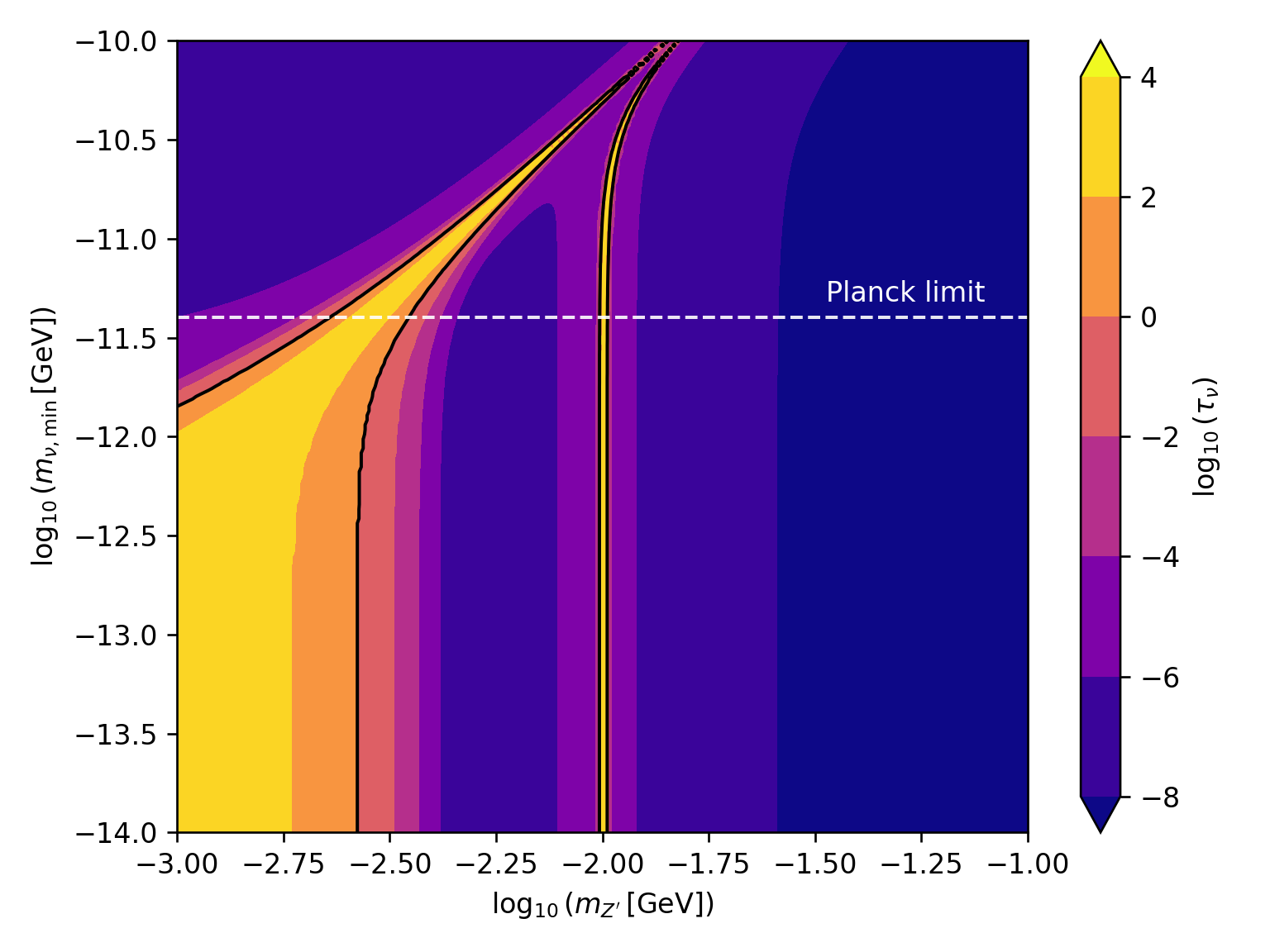}\includegraphics[width=0.6\columnwidth]{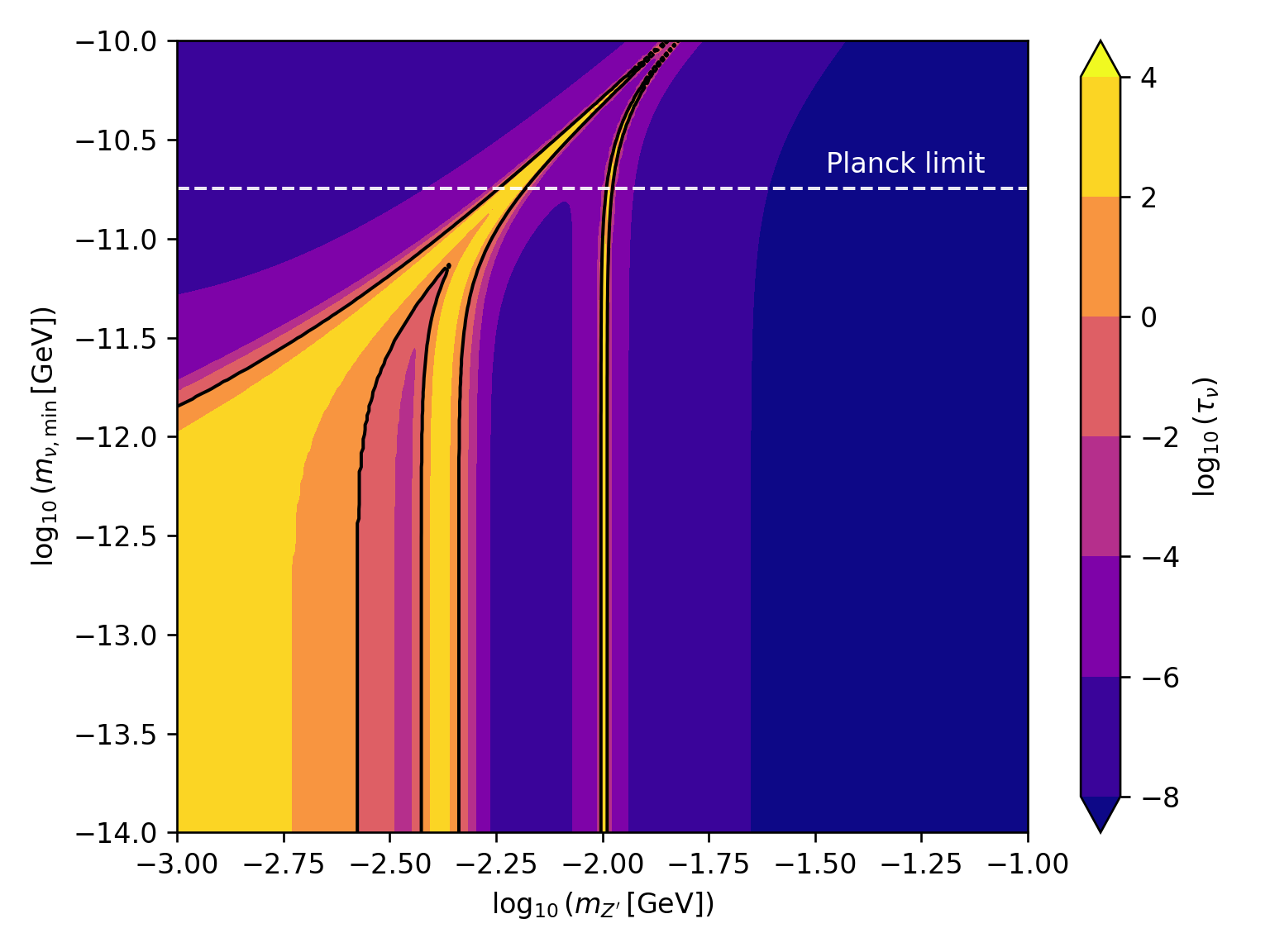}}
    \caption{
   Optical depth $\log_{10}(\tau_\nu)$ as a function of $m_{Z'}$ and $m_\nu$ for the normal (left) and inverted (right) neutrino mass hierarchies. {\color{black} we computed the optical depth for PeV-scale neutrinos ($E_\nu \sim 10^{15}$ eV), which corresponds to the energy range of the highest-energy neutrinos observed by IceCube.} Yellow bands correspond to regions where $\tau_\nu \gtrsim 1$. {\color{black}The regions above the white dashed lines are ruled out by cosmology \cite{Planck:2018vyg}.}
    }
    \label{fig:Z-prime  vs neutrino}
\end{figure}

{\color{black}

A crucial technical point in the optical-depth calculation concerns the
nonzero temperature of the cosmic neutrino background (C$\nu$B).
At the smallest neutrino masses, neglecting the thermal motion of the
background neutrinos leads to an incorrect determination of the
center-of-mass energy and, consequently, of the resonant scattering rate.

Today the C$\nu$B temperature is
$T_{\nu,0} \simeq 1.7 \times 10^{-4}\,\mathrm{eV}$,
corresponding to typical thermal momenta
$p_{\rm th} \sim \mathcal{O}(3 T_{\nu,0})
\sim 5 \times 10^{-4}\,\mathrm{eV}$.
For neutrino masses in the sub-meV regime, the background neutrinos
are therefore semi-relativistic, and their momentum distribution cannot
be approximated by a fixed target at rest.
In this regime, the dominant contribution to the scattering rate arises
from the convolution of the Breit--Wigner resonance structure with the
full Fermi--Dirac momentum distribution.

The thermally averaged cross section for scattering between a
high-energy neutrino of energy $E$ and background neutrinos is
\begin{equation}
\langle \sigma \rangle
=
\frac{2 g_{Z'}^4}{3\pi}\,
I_{\rm thermal}(m_{Z'}, \Gamma_{Z'}, E, T_{\nu,0}),
\end{equation}
where
\begin{equation}
I_{\rm thermal}
=
\frac{
\displaystyle
\int_0^\infty \!\! dp \int_{-1}^{+1} \!\! d\mu \;
\frac{s(p,\mu)}{\left[s(p,\mu)-m_{Z'}^2\right]^2 + m_{Z'}^2 \Gamma_{Z'}^2}
\, f(p)\, p^2
}{
\displaystyle
\int_0^\infty f(p)\, p^2\, dp
}.
\end{equation}
Here
\begin{equation}
s(p,\mu)
=
m_\nu^2 + 2E\!\left(\sqrt{p^2+m_\nu^2}-p\mu\right),
\qquad
f(p)=\frac{1}{e^{p/T_{\nu,0}}+1},
\end{equation}
and $\mu=\cos\theta$ is the scattering angle.
The angular integration can be performed analytically,
while the momentum integral must be evaluated numerically.

The behavior of $I_{\rm thermal}$ is controlled by the relation between
the resonance momentum
\begin{equation}
p_{\rm res} \sim \frac{m_{Z'}^2}{2E}
\end{equation}
and the thermal scale $T_{\nu,0}$.
If $p_{\rm res} \lesssim T_{\nu,0}$, the resonance lies within the bulk
of the thermal distribution and the integral is unsuppressed.
Conversely, when $p_{\rm res} \gg T_{\nu,0}$, only the exponentially
suppressed tail of the Fermi--Dirac distribution can contribute,
and the cross section is strongly damped.

This feature implies that the optical depth does not depend primarily
on the neutrino rest mass at small $m_\nu$, but rather on the ratio
$m_{Z'}^2/(E T_{\nu,0})$.
In particular, naive estimates that replace the momentum integral by
a fixed characteristic value (e.g.\ $p \sim 3T_{\nu,0}$) or that neglect
the angular dependence can misestimate the scattering rate by many
orders of magnitude.
A proper thermal average over both momentum and angle is therefore
essential for a consistent determination of the resonant structure
and of the $\tau_\nu \sim 1$ contours in the parameter space.

}

\begin{figure}
    \centering

\mbox{\hspace*{-1cm}\includegraphics[width=0.6\columnwidth]{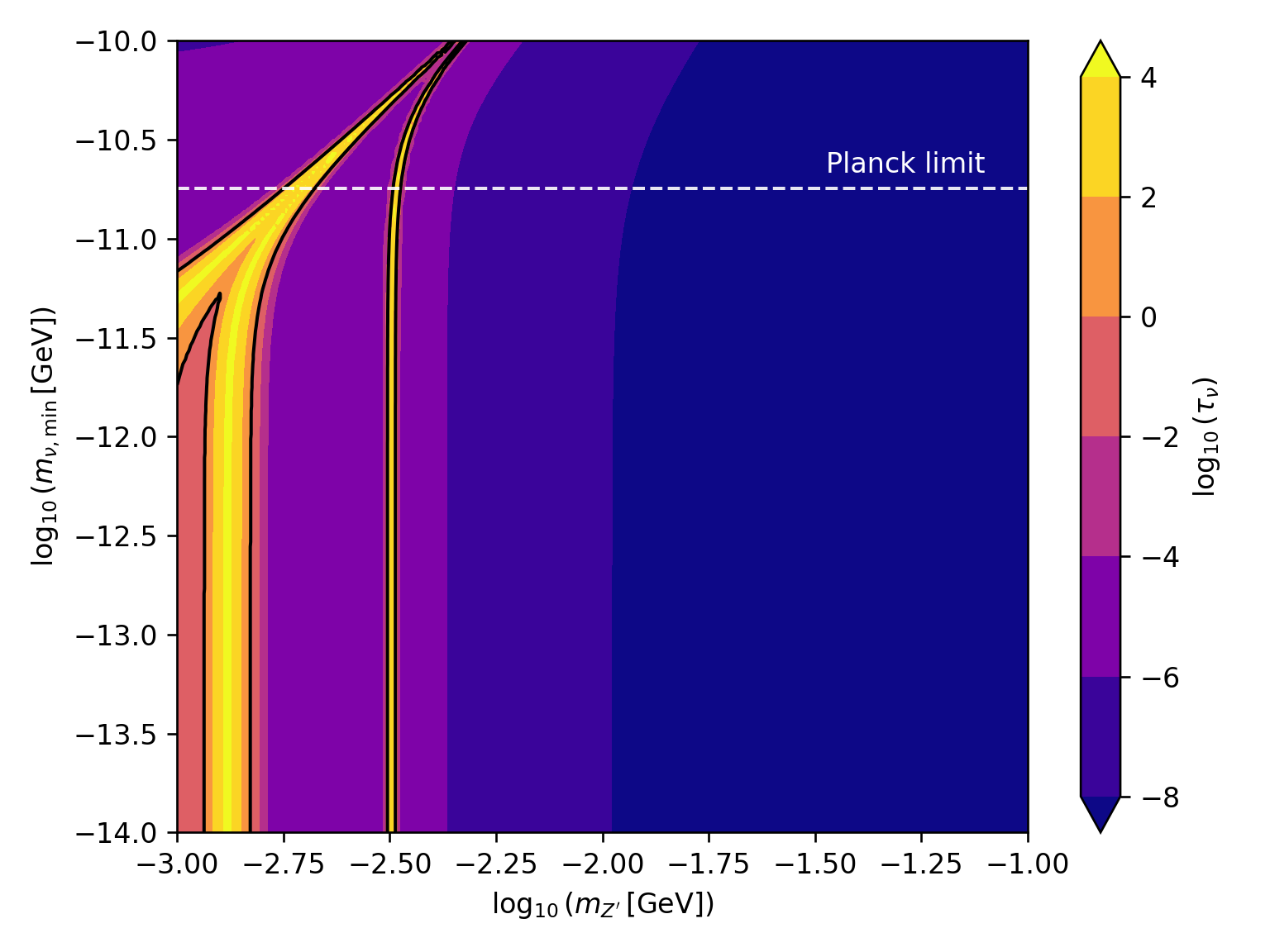}\includegraphics[width=0.6\columnwidth]{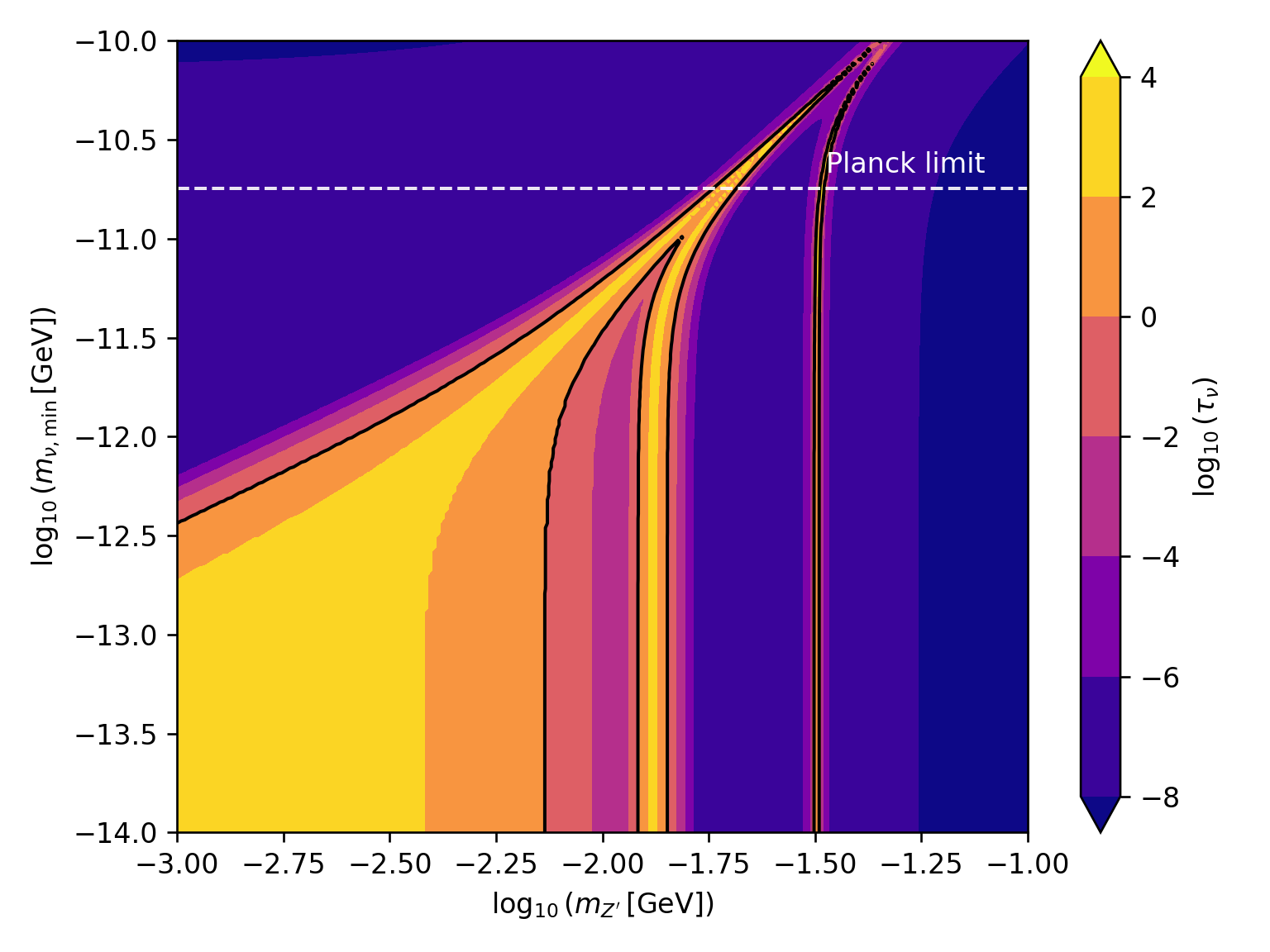}}
    \caption{
   \color{black}As in fig.~\ref{fig:Z-prime  vs neutrino}, but for an impinging neutrino energy of 0.1 PeV (left) and 10 PeV (right), for the normal hierarchy.
    }
    \label{fig:fig2}
\end{figure}

\begin{figure}
    \centering
    \includegraphics[width=0.48\columnwidth]{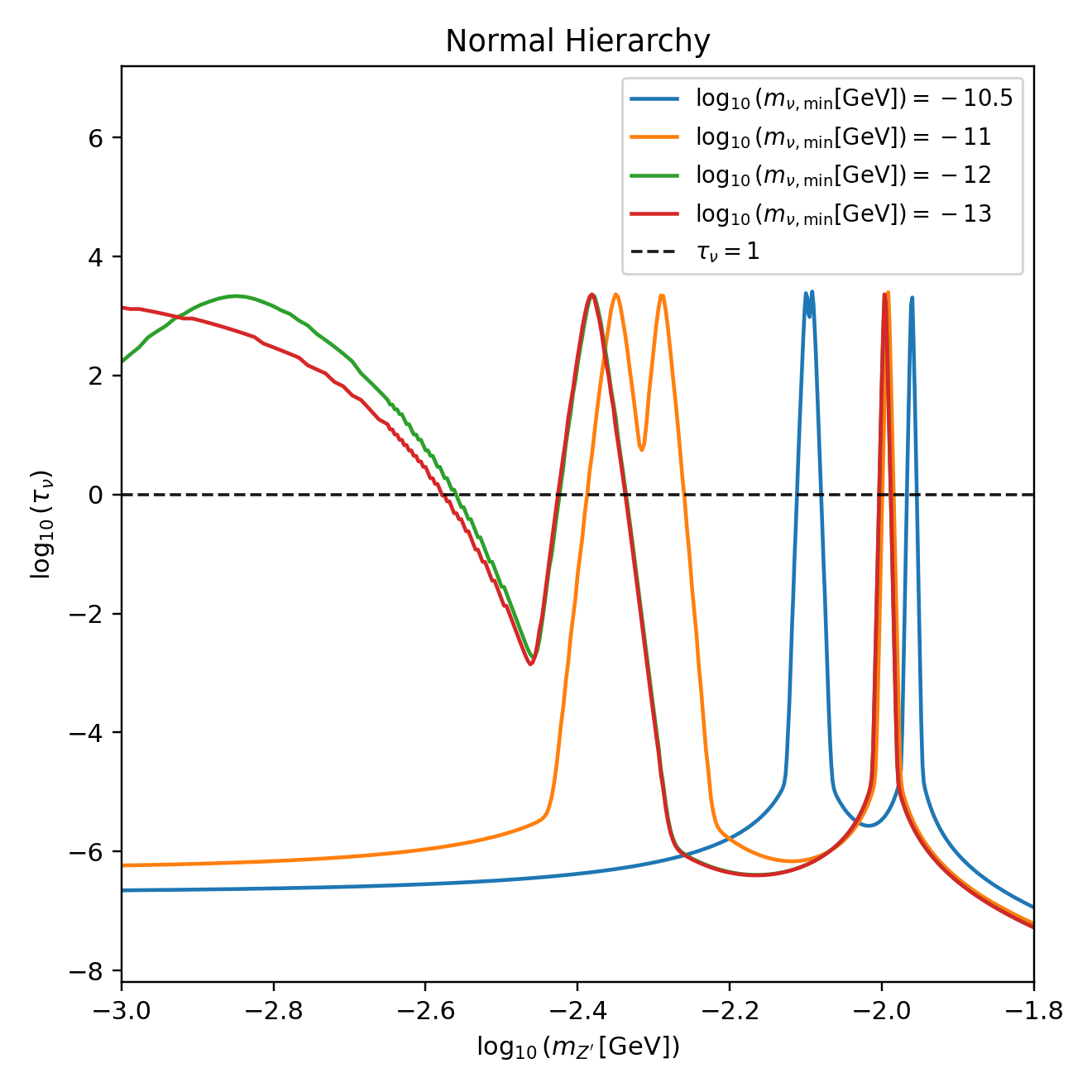}\quad \includegraphics[width=0.48\columnwidth]{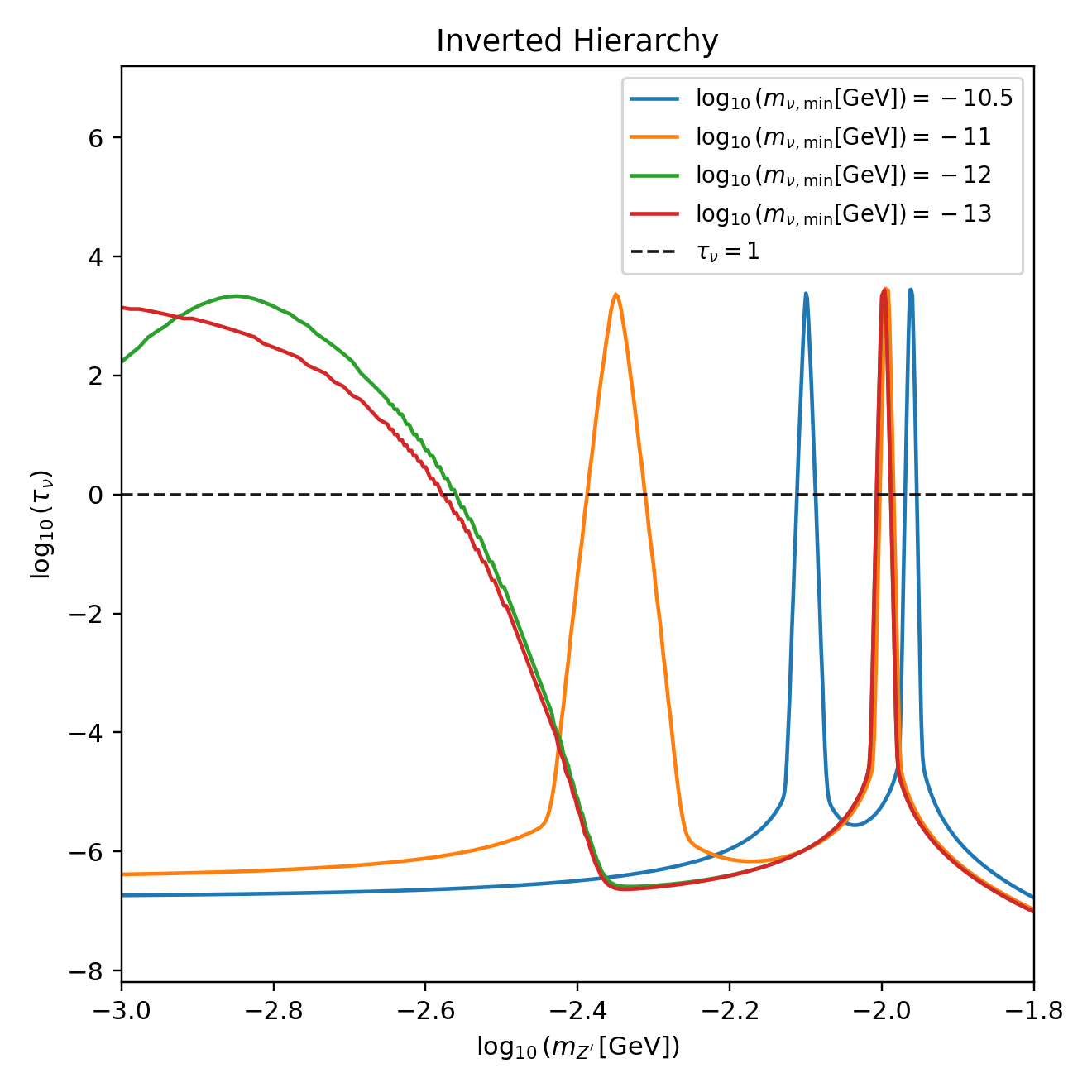}
    \caption{\color{black}
Cutouts of the optical depth $\log_{10}(\tau_\nu)$ as a function of
$\log_{10}(m_{Z'}\,[\mathrm{GeV}])$ for fixed values of the lightest
neutrino mass.
The left panel shows the normal hierarchy and the right panel the
inverted hierarchy.
Colored curves correspond to
$\log_{10}(m_{\nu,\min}[\mathrm{GeV}])=-10.5,-11,-12,$ and $-13$,
as indicated in the legend.
The horizontal dashed line denotes $\tau_\nu = 1$.
Sharp peaks arise when the resonance condition
$m_{Z'}^2 \simeq 2 E p$
is satisfied within the thermally populated region of the cosmic
neutrino background.
For sufficiently small $m_{Z'}$, the resonance momentum lies below
the thermal scale and the optical depth becomes large,
whereas for larger $m_{Z'}$ the required resonance momentum exceeds
the characteristic thermal momentum, leading to exponential
suppression from the tail of the Fermi--Dirac distribution.
The position and width of the peaks therefore reflect the interplay
between the mediator mass, the neutrino energy, and the C$\nu$B
thermal scale rather than a simple dependence on the neutrino
rest mass alone.
}\label{fig:fig3}
\end{figure}

{\color{black}
Figure~\ref{fig:Z-prime  vs neutrino} presents the two–dimensional parameter space in the 
$(\log_{10} m_{Z'}, \log_{10} m_{\nu,\min})$ plane, showing contours of the 
optical depth $\tau_\nu$ computed using the full thermal average over the 
cosmic neutrino background (C$\nu$B). As emphasized above, the relevant 
dynamical scale controlling resonant scattering is not the neutrino rest mass 
itself, but the ratio $m_{Z'}^2/(E T_{\nu,0})$, which determines whether the 
resonance lies within the thermally populated region of the background 
distribution. The white dashed horizontal lines indicate the cosmological 
upper bounds on the lightest neutrino mass for the normal and inverted 
hierarchies, respectively. These bounds remove the upper portions of the 
parameter space in Fig.~\ref{fig:Z-prime  vs neutrino}, leaving only the lower–mass region 
phenomenologically viable. Note that the figure would shift if different neutrino energies were chosen, as we show in Fig.~\ref{fig:fig2} explicitly: For a neutrino energy $E_\nu$, the center-of-mass energy squared is approximately $s \approx 2E_\nu m_\nu$ (for $E_\nu \gg m_\nu$), so the resonance occurs when $m_{Z'} \approx \sqrt{2E_\nu m_\nu}$. Higher energies would shift the resonance bands to higher $Z'$ masses for fixed neutrino mass.

Figure~\ref{fig:fig3} provides one–dimensional cutouts of 
$\log_{10}(\tau_\nu)$ as a function of $m_{Z'}$ for fixed values of the 
lightest neutrino mass, shown separately for the normal and inverted 
hierarchies. These curves make explicit the resonant structure responsible 
for the contours seen in Fig.~\ref{fig:Z-prime  vs neutrino}. For small $m_{Z'}$, the 
resonance occurs at momenta comparable to or below the thermal scale 
$T_{\nu,0}$, leading to large optical depths. As $m_{Z'}$ increases, the 
required resonance momentum $p_{\rm res} \simeq m_{Z'}^2/(2E)$ moves into 
the exponentially suppressed tail of the Fermi–Dirac distribution, and 
$\tau_\nu$ decreases rapidly. The sharp peaks visible in Fig.~\ref{fig:fig3} 
correspond to the narrow range where the resonance condition overlaps with 
the thermally populated region. 

Importantly, the comparison between Figs.~\ref{fig:Z-prime  vs neutrino} and 
\ref{fig:fig3} demonstrates that the optical depth is controlled, at small masses, by the 
thermal distribution rather than by a naive scaling with the neutrino mass. 
While $m_\nu$ still enters through the kinematics and the hierarchy–dependent 
mass splittings, its role is secondary once the full thermal averaging is 
implemented. The dominant physical effect is instead the interplay between 
the mediator mass, the neutrino energy, and the C$\nu$B temperature. 
This clarifies the origin of the resonant structure and resolves the 
concern raised regarding the treatment of the background temperature.
}

We compare these results to constraints and preferred regions from other cosmological and particle physics observations. In particular, we compute the gauge coupling $g_{Z'}$ required to explain the observed muon anomalous magnetic moment $\Delta a_\mu$ as a function of $m_{Z'}$ using:
\begin{equation}
    \Delta a_\mu(m_{Z'}, g_{Z'}) = 
    \frac{g_{Z'}^2 m_\mu^2}{4\pi^2 m_{Z'}^2} 
    \int_0^1 \frac{x^2 (1-x) \, dx}{1 - x + \left( \frac{m_\mu^2}{m_{Z'}^2} \right) x^2} \,,
    \label{eq:muon_magnetic_moment}
\end{equation}
which can be inverted to express the required coupling:
\begin{equation}
    g_{Z'}(m_{Z'}, \Delta a_\mu) = 
    \left(
        \frac{m_\mu^2}
             {4\pi^2 m_{Z'}^2 \Delta a_\mu}
        \int_0^1 
        \frac{x^2 (1-x) \, dx}
            {1 - x + \left( \frac{m_\mu^2}{m_{Z'}^2} \right) x^2}
    \right)^{-1/2} \,.
    \label{eq:gauge_couple_of_muon_magnetic_moment}
\end{equation}

\begin{figure}
    \centering
    \includegraphics[width=0.48\columnwidth]{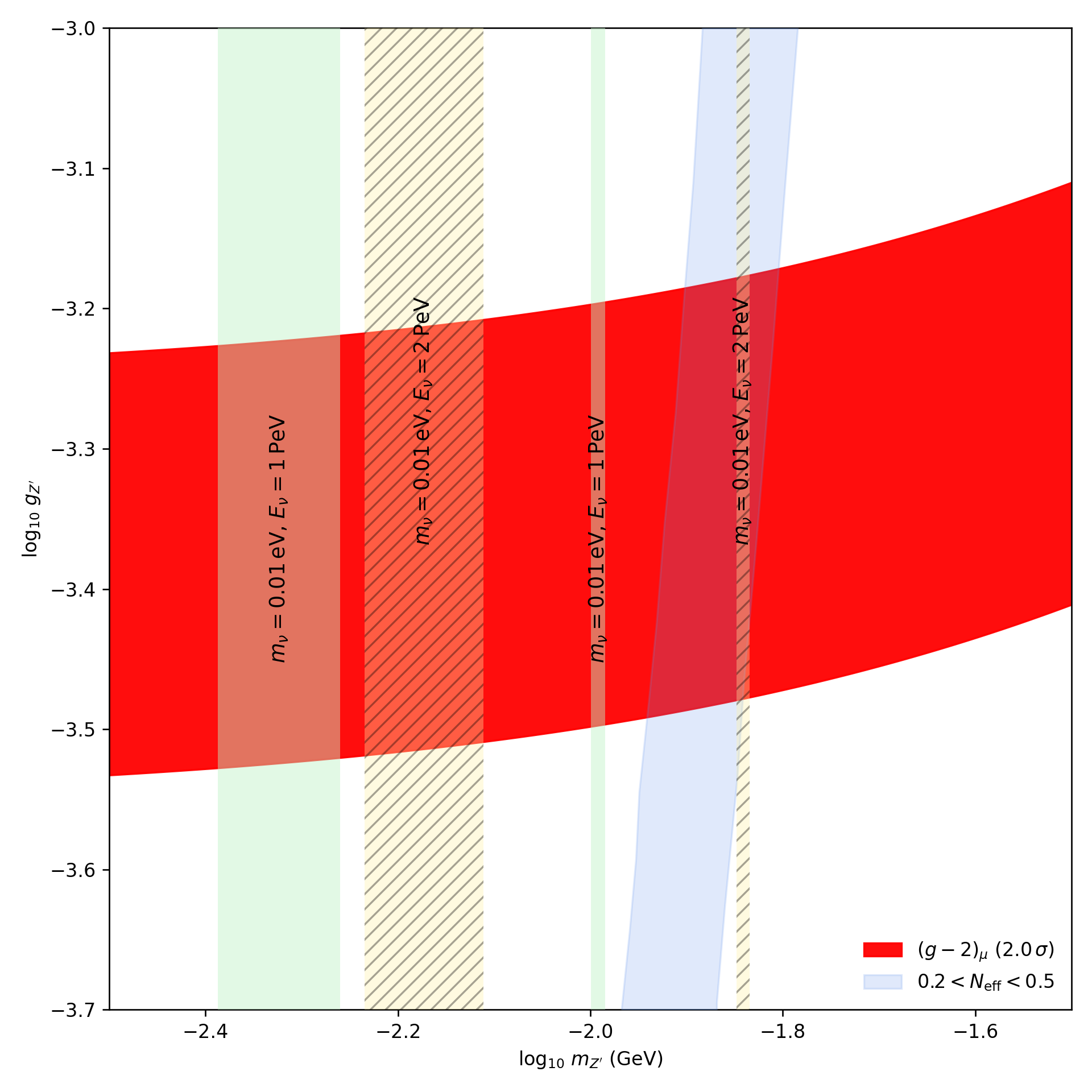}\quad\includegraphics[width=0.48\columnwidth]{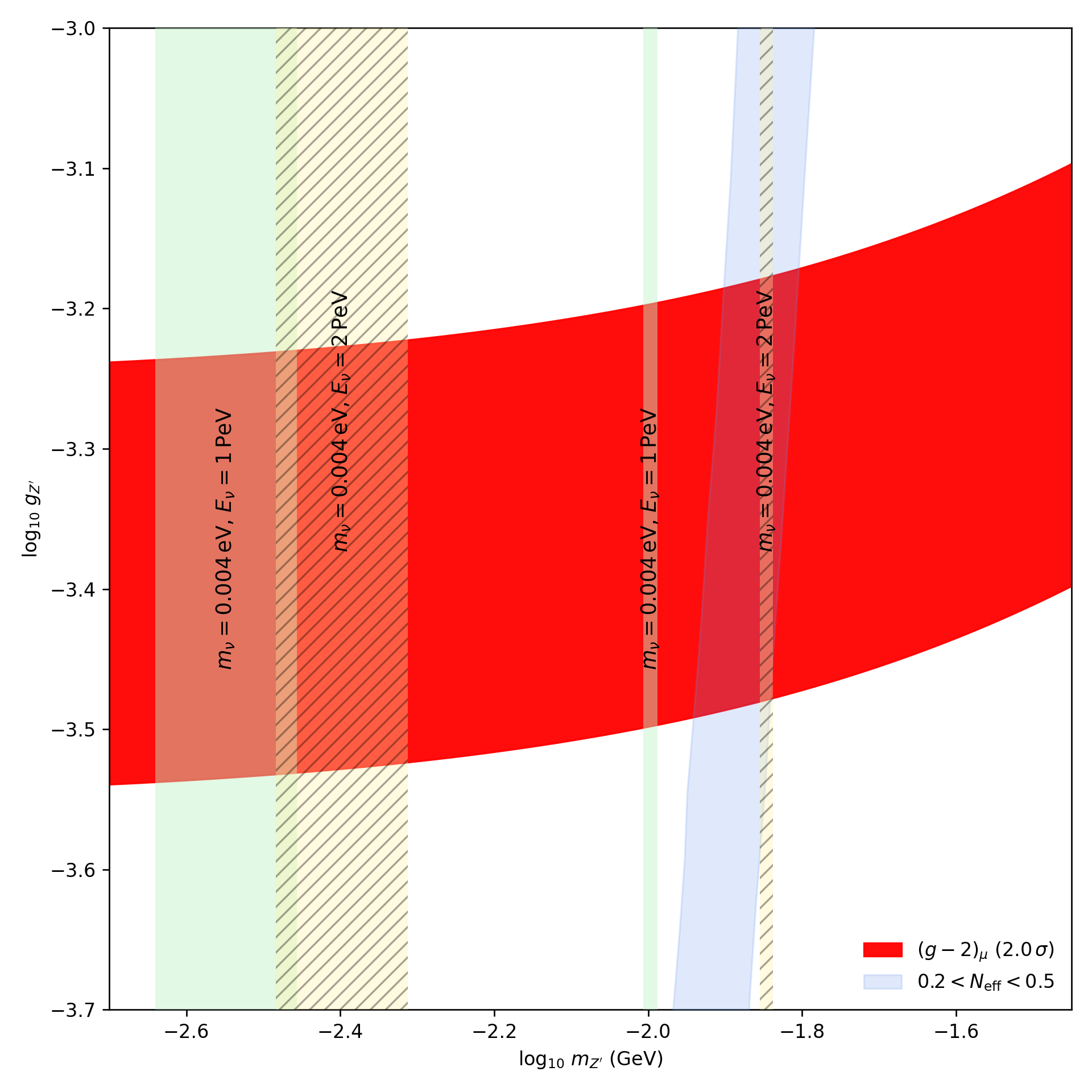}
   \caption{\color{black}
Parameter space in the $(\log_{10} m_{Z'}, \log_{10} g_{Z'})$ plane showing 
the $(g-2)_\mu$ $2\sigma$ preferred region (red) and the region where 
$0.2 < \Delta N_{\rm eff} < 0.5$ (light blue). 
Superimposed vertical bands indicate the ranges of mediator mass $m_{Z'}$ 
for which the optical depth satisfies $\tau_\nu > 1$ for selected values 
of the lightest neutrino mass and representative neutrino energies. 
The band widths reflect the resonance condition and the full thermal 
averaging over the cosmic neutrino background. 
The overlap (or lack thereof) between these bands and the $(g-2)_\mu$ 
region illustrates the degree to which efficient high-energy neutrino 
attenuation can occur within phenomenologically viable couplings.}

    \label{fig:muong2_vs_Neff}
\end{figure}

{\color{black}

Figure~\ref{fig:muong2_vs_Neff} synthesizes the various constraints in the 
$(\log_{10} m_{Z'}, \log_{10} g_{Z'})$ plane and illustrates explicitly the 
interplay between neutrino attenuation, the $(g-2)_\mu$ anomaly, and the 
$\Delta N_{\rm eff}$ region. The red band denotes the $(g-2)_\mu$ preferred 
region at $2\sigma$, obtained by inverting the measured $\Delta a_\mu$ for 
each value of $m_{Z'}$. The light blue band corresponds to the region where 
$0.2 < \Delta N_{\rm eff} < 0.5$, which is of interest for alleviating the 
Hubble tension.

Superimposed on these regions are vertical bands indicating the ranges of 
$m_{Z'}$ for which the optical depth satisfies $\tau_\nu > 1$ for selected 
values of the lightest neutrino mass and representative neutrino energies, respectively 1 and 2 PeV. 
The bands are vertical because, once $g_{Z'}(m_{Z'})$ is fixed by the 
$(g-2)_\mu$ inversion, the optical depth condition effectively selects a 
restricted interval in mediator mass. The width of each band reflects the 
resonant condition together with the thermal averaging over the cosmic 
neutrino background.

Several features are apparent. First, for $E_\nu \sim \mathcal{O}({\rm PeV})$, 
the $\tau_\nu>1$ bands lie at mediator masses near 
$m_{Z'} \sim 10^{-2}\,\mathrm{GeV}$, corresponding to resonance momenta 
$p_{\rm res} \simeq m_{Z'}^2/(2E_\nu)$ that fall within the thermally populated 
region of the background distribution. As $E_\nu$ increases, the resonance 
condition shifts to larger $m_{Z'}$, and the corresponding vertical bands 
move accordingly. 

Second, the overlap between the $\tau_\nu>1$ bands and the red $(g-2)_\mu$ 
region is highly sensitive to both the neutrino mass and the neutrino energy. 
For some combinations, the attenuation region partially intersects the 
$(g-2)_\mu$ preferred band, implying that efficient high-energy neutrino 
absorption can occur within phenomenologically viable couplings. In other 
cases, the $\tau_\nu>1$ bands lie entirely outside the red region, indicating 
that significant attenuation would require couplings incompatible with the 
muon anomalous magnetic moment constraint.

Third, the comparison with the light blue $\Delta N_{\rm eff}$ region shows 
that the parameter space capable of modifying early-Universe radiation 
content does not generically coincide with the region where high-energy 
neutrino attenuation is efficient. The degree of overlap depends on the 
hierarchy and on the assumed neutrino mass. This demonstrates that the 
cosmological and astrophysical effects probe complementary, and in some 
cases disjoint, regions of parameter space.

Overall, Fig.~\ref{fig:muong2_vs_Neff} highlights the nontrivial interplay between 
laboratory constraints, cosmological considerations, and high-energy 
astrophysical observables. The vertical structure of the attenuation bands 
reflects the thermal origin of the resonance, controlled primarily by the 
ratio $m_{Z'}^2/(E_\nu T_{\nu,0})$, rather than by the neutrino rest mass 
alone. This reinforces the importance of the full thermal averaging 
procedure in determining the viable regions of the model.

}

{\color{black}

\section{Discussion and Conclusions}

In this work we have revisited resonant neutrino scattering in a 
$U(1)_{L_\mu-L_\tau}$ framework, focusing on its implications for the 
optical depth of high-energy neutrinos propagating through the cosmic 
neutrino background (C$\nu$B). While similar mediator scenarios have been 
studied previously (e.g., Ref.~\cite{Carpio:2021jhu}), our analysis introduces several 
technical refinements and presents the results in a framework that 
clarifies the role of absolute neutrino mass and thermal effects.

The central methodological improvement of this work is the implementation 
of a fully thermal-averaged cross section. Instead of approximating the 
center-of-mass energy using $s \simeq 2 m_\nu E_\nu$, we perform the 
complete integration over the Fermi--Dirac momentum distribution and the 
angular dependence of the background neutrinos. This correction becomes 
essential when the neutrino rest mass is comparable to or smaller than the 
thermal momentum scale $T_{\nu,0}$. In this regime, the resonance structure 
is governed not by $m_\nu$ alone but by the ratio 
$m_{Z'}^2/(E_\nu T_{\nu,0})$, and the cross section can be exponentially 
suppressed when the required resonance momentum lies in the high-momentum 
tail of the distribution. 

As a result, the optical depth at small $m_\nu$ behaves qualitatively 
differently from naive treatments that neglect thermal motion. This 
modifies both the scaling of $\tau_\nu$ and the interpretation of the 
allowed parameter space. In particular, we find that the resonance 
structure in Fig.~1 is controlled by thermal kinematics at small mass, 
while at larger $m_\nu$ the conventional rest-mass scaling is recovered.

Our presentation emphasizes the optical depth $\tau_\nu$ itself as the 
primary observable quantity. Figure~\ref{fig:Z-prime  vs neutrino} maps $\tau_\nu$ in the 
$(\log_{10} m_{Z'}, \log_{10} m_{\nu,\min})$ plane for fixed energies, 
revealing structure that is not immediately visible in traditional 
$(m_{Z'}, g_{Z'})$ plots. Figures~\ref{fig:fig2} and~\ref{fig:fig3} provide complementary 
energy-resolved visualizations and cutouts, illustrating how the 
resonance position and width shift with neutrino energy and absolute 
mass scale. The comparison between normal and inverted hierarchies shows 
that the hierarchy dependence is modest at small masses once thermal 
averaging is included, but becomes more pronounced at larger $m_\nu$.

Figure~\ref{fig:muong2_vs_Neff} synthesizes these ingredients by overlaying the regions 
favored by $(g-2)_\mu$, those consistent with 
$0.2 < \Delta N_{\rm eff} < 0.5$, and the bands where 
$\tau_\nu > 1$ for representative neutrino masses and energies. 
We find that non-empty regions of parameter space exist where all 
three conditions can be simultaneously satisfied. However, the 
existence and location of such regions depend sensitively on both 
the lightest neutrino mass and the energy of the propagating 
neutrino. For smaller lightest neutrino masses $m_{\nu}$, thermal suppression shifts 
or narrows the $\tau_\nu > 1$ bands, while increasing the neutrino 
energy moves the resonance condition and can restore overlap with 
the cosmologically and $(g-2)_\mu$ preferred regions. 

In this sense, the viability of a significant cosmic neutrino opacity 
is not a universal prediction of the model, but rather a conditional 
statement tied to the absolute neutrino mass scale and the energy 
spectrum of astrophysical neutrinos. The parameter space compatible 
with $(g-2)_\mu$, $\Delta N_{\rm eff}$, and $\tau_\nu>1$ is therefore 
structured and energy-dependent, rather than generic.

We have also incorporated updated cosmological limits on the sum of 
neutrino masses and made explicit how they intersect the optical-depth 
structure. These constraints further restrict the allowed parameter 
space in a hierarchy-dependent manner.

Taken together, our results demonstrate that a consistent thermal 
treatment of the C$\nu$B is essential for interpreting resonant 
neutrino scattering in the sub-eV mass regime. While the underlying 
mediator framework overlaps with earlier work, the combination of 
(i) a complete thermal average, 
(ii) a systematic mapping of optical depth as a function of absolute 
neutrino mass, 
(iii) explicit hierarchy comparisons, 
(iv) energy-resolved analyses, and 
(v) updated cosmological overlays 
yields quantitatively distinct conclusions and a clearer picture of 
the physically viable parameter space.

Future improvements in measurements of the absolute neutrino mass 
scale, together with high-statistics observations of PeV-scale 
astrophysical neutrinos, will be essential in further testing this 
scenario and refining the interplay between particle physics, 
cosmology, and neutrino astronomy.

}

\acknowledgments
%%%%%%%%%%%%%%%%%%%%%%%%%%%%%%%%%%%%%%%%%%%%%%%%%%%%%
%%%%%%%%%%%%%%%%%%%%%%%%%%%%%%%%%%%%%%%%%%%%%%%%%%%%%
This work is partly supported by the U.S.\ Department of Energy grant number de-sc0010107 (SP).

% Bibliography

%% [A] Recommended: using JHEP.bst file
\bibliographystyle{JHEP}

\bibliography{biblio.bib}

@article{Hooper_2023,
  title = {Signals of a new gauge boson from IceCube and the muon $g\ensuremath{-}2$},
  author = {Hooper, Dan and Juan, Joaquim Iguaz and Serpico, Pasquale D.},
  journal = {Phys. Rev. D},
  volume = {108},
  issue = {2},
  pages = {023007},
  numpages = {14},
  year = {2023},
  month = {Jul},
  publisher = {American Physical Society},
  doi = {10.1103/PhysRevD.108.023007},
  url = {https://link.aps.org/doi/10.1103/PhysRevD.108.023007}
}

@article{Escudero_2019,
   title={Cosmology with a very light $L_\mu-L_\tau$ gauge boson},
   volume={2019},
   ISSN={1029-8479},
   url={http://dx.doi.org/10.1007/JHEP03(2019)071},
   DOI={10.1007/jhep03(2019)071},
   number={3},
   journal={Journal of High Energy Physics},
   publisher={Springer Science and Business Media LLC},
   author={Escudero, Miguel and Hooper, Dan and Krnjaic, Gordan and Pierre, Mathias},
   year={2019},
   month=mar }

@article{Carpio:2021jhu,
    author = "Carpio, Jose Alonso and Murase, Kohta and Shoemaker, Ian M. and Tabrizi, Zahra",
    title = "{High-energy cosmic neutrinos as a probe of the vector mediator scenario in light of the muon g-2 anomaly and Hubble tension}",
    eprint = "2104.15136",
    archivePrefix = "arXiv",
    primaryClass = "hep-ph",
    doi = "10.1103/PhysRevD.107.103057",
    journal = "Phys. Rev. D",
    volume = "107",
    number = "10",
    pages = "103057",
    year = "2023"
}

@article{Fang:2024abc,
  author = {Rundong Fang and Ji-Heng Guo and Jia Liu and Xiao-Ping Wang},
  title = {{Muon g-2, Long-Range Muon Spin Force, and Neutrino Oscillations}},
  journal = {Phys. Rev. D},
  volume = {110},
  pages = {035037},
  year = {2024},
  doi = {10.1103/PhysRevD.110.035037},
  eprint = {2405.02084},
  archivePrefix = {arXiv},
  primaryClass = {hep-ph}
}

@article{Muong-2:2023cdq,
  author = {{Muon g-2 Collaboration}},
  title = {{Measurement of the positive muon anomalous magnetic moment to 0.20 ppm}},
  journal = {Phys. Rev. Lett.},
  volume = {131},
  pages = {161802},
  year = {2023},
  doi = {10.1103/PhysRevLett.131.161802},
  eprint = {2308.06230},
  archivePrefix = {arXiv},
  primaryClass = {hep-ex}
}

@article{Altmannshofer:2014pba,
  author = {Wolfgang Altmannshofer and Stefania Gori and Maxim Pospelov and Itay Yavin},
  title = {{Neutrino Trident Production: A Powerful Probe of New Physics with Neutrino Beams}},
  journal = {Phys. Rev. Lett.},
  volume = {113},
  pages = {091801},
  year = {2014},
  doi = {10.1103/PhysRevLett.113.091801},
  eprint = {1406.2332},
  archivePrefix = {arXiv},
  primaryClass = {hep-ph}
}

@article{Gould:1966paz,
  author = {R. J. Gould and G. P. Schréder},
  title = {{Opacity of the Universe to High-Energy Photons}},
  journal = {Phys. Rev. Lett.},
  volume = {16},
  pages = {252--254},
  year = {1966},
  doi = {10.1103/PhysRevLett.16.252}
}

@article{Roulet:1993pz,
  author = {Esteban Roulet},
  title = {{Ultrahigh-energy neutrino absorption by neutrino dark matter}},
  journal = {Phys. Rev. D},
  volume = {47},
  pages = {5247--5252},
  year = {1993},
  doi = {10.1103/PhysRevD.47.5247}
}

@article{Yoshida:1996ie,
  author = {Shigeru Yoshida and Masataka Teshima},
  title = {{Energy spectrum of ultrahigh-energy cosmic rays with extragalactic origin}},
  journal = {Prog. Theor. Phys.},
  volume = {89},
  pages = {833--845},
  year = {1993},
  doi = {10.1143/PTP.89.833}
}

@article{Ng:2014pca,
  author = {Kenny C. Y. Ng and John F. Beacom},
  title = {{Cosmic neutrino cascades from secret neutrino interactions}},
  journal = {Phys. Rev. D},
  volume = {90},
  pages = {065035},
  year = {2014},
  doi = {10.1103/PhysRevD.90.065035},
  eprint = {1404.2288},
  archivePrefix = {arXiv},
  primaryClass = {astro-ph.HE}
}

@article{Ioka:2014kca,
  author = {Kunihito Ioka and Kohta Murase},
  title = {{IceCube PeV–EeV neutrinos and secret interactions of neutrinos}},
  journal = {PTEP},
  volume = {2014},
  pages = {061E01},
  year = {2014},
  doi = {10.1093/ptep/ptu090},
  eprint = {1404.2279},
  archivePrefix = {arXiv},
  primaryClass = {astro-ph.HE}
}

@article{Esteban:2020cvm,
  author = {I. Esteban and M. C. Gonzalez-Garcia and M. Maltoni and T. Schwetz and A. Zhou},
  title = {{The fate of hints: updated global analysis of three-flavor neutrino oscillations}},
  journal = {JHEP},
  volume = {09},
  pages = {178},
  year = {2020},
  doi = {10.1007/JHEP09(2020)178},
  eprint = {2007.14792},
  archivePrefix = {arXiv},
  primaryClass = {hep-ph}
}

@article{KATRIN:2021uub,
  author = {{KATRIN Collaboration}},
  title = {{Direct neutrino-mass measurement with sub-electronvolt sensitivity}},
  journal = {Nature Phys.},
  volume = {18},
  pages = {160--166},
  year = {2022},
  doi = {10.1038/s41567-021-01463-1},
  eprint = {2105.08533},
  archivePrefix = {arXiv},
  primaryClass = {hep-ex}
}

@article{Planck:2018vyg,
  author = {{Planck Collaboration}},
  title = {{Planck 2018 results. VI. Cosmological parameters}},
  journal = {Astron. Astrophys.},
  volume = {641},
  pages = {A6},
  year = {2020},
  doi = {10.1051/0004-6361/201833910},
  eprint = {1807.06209},
  archivePrefix = {arXiv},
  primaryClass = {astro-ph.CO}
}

@article{Minkowski:1977sc,
  author = {P. Minkowski},
  title = "{{\ensuremath{\mu}}{\textrightarrow}e {\ensuremath{\gamma}} at a rate of one out of 10 $^{9}$ muon decays?}",
  journal = {Phys. Lett. B},
  volume = {67},
  pages = {421--428},
  year = {1977},
  doi = {10.1016/0370-2693(77)90435-X}
}

@article{Mohapatra:1980yp,
  author = {R. N. Mohapatra and G. Senjanovic},
  title = {{Neutrino Mass and Spontaneous Parity Violation}},
  journal = {Phys. Rev. Lett.},
  volume = {44},
  pages = {912},
  year = {1980},
  doi = {10.1103/PhysRevLett.44.912}
}

@article{Muong-2:2025final,
  author       = {{Muon g-2 Collaboration}},
  title        = {{Final Combined Measurement of the Muon Anomalous Magnetic Moment at Fermilab}},
  journal      = {Phys.\ Rev.\ Lett.},
  year         = {2025},
  eprint       = {2506.03069},
  archivePrefix= {arXiv},
  primaryClass = {hep-ex},
  submitted    = {2 June 2025},
  note         = {Submitted June 2025; Final combined result with 127 ppb precision}
}

@article{Mainz:2025HVP,
  author       = {Meyer, N. and Miller, K. and Ottnad, K. and Parrino, J. and Risch, A. and Wittig, H.},
  title        = {Lattice QCD Determination of the Long–Distance Contribution to the Hadronic Vacuum Polarization for Muon \(g-2\)},
  journal      = {JHEP},
  volume       = {04},
  pages        = {098},
  year         = {2025},
  doi          = {10.1007/JHEP04(2025)098},
}

@article{Davier:2024piipi,
    author = "Davier, Michel and Hoecker, Andreas and Lutz, Anne-Marie and Malaescu, Bogdan and Zhang, Zhiqing",
    title = "{Tensions in $e^+e^-\rightarrow \pi ^+\pi ^-(\gamma )$ measurements: the new landscape of data-driven hadronic vacuum polarization predictions for the muon $g - 2$}",
    eprint = "2312.02053",
    archivePrefix = "arXiv",
    primaryClass = "hep-ph",
    doi = "10.1140/epjc/s10052-024-12964-7",
    journal = "Eur. Phys. J. C",
    volume = "84",
    number = "7",
    pages = "721",
    year = "2024"
}

@article{WhitePaper:2025update,
  author       = {{Muon g-2 Theory Initiative}},
  title        = {{The anomalous magnetic moment of the muon in the Standard Model: an update}},
  journal      = {Phys.\ Rept.},
  year         = {2025},
  note         = {Revision including new CMD‑3 data},
  eprint       = {2505.21476},
}

@article{LatticeReview:2025,
  author       = {Fodor, Z. and Gérardin, A. and others},
  title        = {Recent Progress in Muon \(g-2\) from Lattice QCD},
  journal      = {Proceedings of LatticeNET 2025},
  year         = {2025},
  note         = {See slides "Recent progress in the study of the muon \(g-2\) from lattice QCD"},
}

@article{Holst:2022predictive,
  author = {I. Holst and D. Hooper and G. Krnjaic},
  title = {Simplest and Most Predictive Model of Muon $g-2$ and Thermal Dark Matter},
  journal = {Phys. Rev. Lett.},
  volume = {128},
  year = {2022},
  pages = {141802},
  doi = {10.1103/PhysRevLett.128.141802},
  eprint = {2107.09067},
}

@article{IceCube:2013dk,
  author = {{IceCube Collaboration}},
  title = {{Evidence for High-Energy Extraterrestrial Neutrinos at the IceCube Detector}},
  journal = {Science},
  volume = {342},
  year = {2013},
  pages = {1242856},
  doi = {10.1126/science.1242856},
}
\end{document}